\documentclass[
final
]{dmtcs-episciences}

\usepackage[utf8]{inputenc}
\usepackage{subfigure}

\usepackage[round]{natbib}

\usepackage{amsmath,amssymb,amsfonts,amsthm,latexsym}
\usepackage{cleveref}
\usepackage{xspace}
\usepackage{mathtools}

\makeatletter
\def\moverlay{\mathpalette\mov@rlay}
\def\mov@rlay#1#2{\leavevmode\vtop{%
   \baselineskip\z@skip \lineskiplimit-\maxdimen
   \ialign{\hfil$\m@th#1##$\hfil\cr#2\crcr}}}
\newcommand{\charfusion}[3][\mathord]{
    #1{\ifx#1\mathop\vphantom{#2}\fi
        \mathpalette\mov@rlay{#2\cr#3}
      }
    \ifx#1\mathop\expandafter\displaylimits\fi}
\makeatother
\newcommand{\cupdot}{\charfusion[\mathbin]{\cup}{\cdot}}

\newcommand{\mc}{\mathcal}
\newcommand{\lca}{\ensuremath{\operatorname{\mathrm{lca}}}}
\newcommand{\lc}[1]{\ensuremath{_{{\scriptscriptstyle #1}}}}
\newcommand{\NP}{\textsf{NP}}
\newcommand{\irr}[1]{\ensuremath{[#1]_{\textrm{irr}}}\xspace}
\newcommand{\X}{\ensuremath{X}\xspace}
\newcommand{\PS}[1]{\ensuremath{\mc{P}(#1)}\xspace}
\newcommand{\Q}{\ensuremath{\mc{Q}}\xspace}
\newcommand{\eps}{\ensuremath{\varepsilon}\xspace}
\newcommand{\eXM}{\ensuremath{\eps \colon \irr{\X \times \X} \to \PS{M}}\xspace}
\newcommand{\G}{\ensuremath{\mc{G}}\xspace}
\newcommand{\NnotCol}[1]{\ensuremath{{N}_{\neg #1}}\xspace}
\newcommand{\Ns}[1]{\ensuremath{\mc{N}\ifthenelse{\equal{#1}{}}{}{_{\neg #1}}}\xspace}
\newcommand{\diam}{\mathrm{diam}\xspace}
\newcommand{\Ss}{\ensuremath{\mc{S}}\xspace}
\newcommand{\SsS}{\ensuremath{\mc{S}^{\star}}\xspace}
\newcommand{\Teps}{\ensuremath{T_{\eps}}\xspace}
\newcommand{\leps}{\ensuremath{\lambda_{\eps}}\xspace}

\crefname{theorem}{Thm.}{Thm.}
\Crefname{theorem}{Theorem}{Theorems}
\crefname{lemma}{L.}{L.}
\Crefname{lemma}{Lemma}{Lemmas}
\crefname{proposition}{Prop.}{Prop.}
\Crefname{proposition}{Proposition}{Propositions}
\crefname{corollary}{Cor.}{Cor.}
\Crefname{corollary}{Corollary}{Corollaries}
\crefname{definition}{Def.}{Def.}
\Crefname{definition}{Definition}{Definitions}
\crefname{figure}{Fig.}{Fig.}
\Crefname{figure}{Figure}{Figures}
\crefname{algorithm}{Alg.}{Alg.}
\Crefname{algorithm}{Algorithm}{Algorithms}

\newtheorem{theorem}{Theorem}[section]
\newtheorem{lemma}[theorem]{Lemma}
\newtheorem{proposition}[theorem]{Proposition}
\newtheorem{corollary}[theorem]{Corollary}
\theoremstyle{definition}
\newtheorem{definition}[theorem]{Definition}
\newtheorem*{remark}{Remark}
\newtheorem*{problem}{Problem}
\newcommand{\PROBLEM}[1]{{\sc #1}}

\author{Marc Hellmuth\affiliationmark{1}
  \and Carsten R.\ Seemann\affiliationmark{2,3}
  \and Peter F.\ Stadler\affiliationmark{2-6}%
  \thanks{Supported in part by the
    \emph{Deutsche Forschungsgemeinschaft grant STA 850/49-1}}}
\title[Generalized Fitch Graphs III]{Generalized Fitch Graphs III: Symmetrized Fitch maps and Sets
    of Symmetric Binary Relations that are explained by Unrooted
    Edge-labeled Trees}
\affiliation{
  Department of Mathematics, Stockholm University, Sweden  \\
  Max Planck Institute for Mathematics in the Sciences, Leipzig, Germany \\
  Bioinformatics Groups, Leipzig University, Germany\\
  Institute for Theoretical Chemistry, University of Vienna, Wien, Austria\\
  Facultad de Ciencias, Universidad Nacional de Colombia, Bogot{\'a}, Colombia\\
  The Santa Fe Institute, Santa Fe, United States}
\received{2020-01-21}
\revised{2021-01-21}
\accepted{2021-05-10}
\begin{document}
\publicationdetails{23}{2021}{1}{13}{6040}
\maketitle
\begin{abstract}
  Binary relations derived from labeled rooted trees play an import role in
  mathematical biology as formal models of evolutionary relationships. The
  (symmetrized) Fitch relation formalizes xenology as the pairs of genes
  separated by at least one horizontal transfer event.  As a natural
  generalization, we consider symmetrized Fitch maps, that is, symmetric
  maps $\varepsilon$ that assign a subset of colors to each pair of
  vertices in $X$ and that can be explained by a tree $T$ with edges that
  are labeled with subsets of colors in the sense that the color $m$
  appears in $\varepsilon(x,y)$ if and only if $m$ appears in a label along
  the unique path between $x$ and $y$ in $T$.  We first give an alternative
  characterization of the monochromatic case and then give a
  characterization of symmetrized Fitch maps in terms of compatibility of a
  certain set of quartets. We show that recognition of symmetrized Fitch
  maps is NP-complete. In the restricted case where
  $|\varepsilon(x,y)|\leq 1$ the problem becomes polynomial, since such
  maps coincide with class of monochromatic Fitch maps whose
  graph-representations form precisely the class of complete multi-partite
  graphs.
\end{abstract}

\sloppy

\section{Introduction}
\label{sec:intro}
Labeled phylogenetic trees are a natural structure to model evolutionary
histories in biology. The leaf set $L$ of the tree $T$ correspond to
currently living entities, while inner nodes model the branching of
lineages that then evolve independently. Labels on vertices and edges
annotate further details on evolutionary events. Considering the evolution
of gene families, for instance, vertex labels may be used to distinguish
gene duplication events from speciation and horizontal gene transfer
\cite{Fitch:00}.  Edge labels, on the other hand, may be used to designate
(rare) events that change properties of genes, genomes, and organisms
\cite{Hellmuth:18b} or to distinguish different fates of offspring genes
such as the horizontal transfer into another genomes
\cite{Geiss:18a}. Distance-based phylogenetics can be seen as special case
of the latter setting, where edges are weighted by evolutionary distances
\cite{Semple2003}. Relations on $L$ are naturally defined as functions of
the edge and/or vertex labels along the unique path connecting a pair of
leaves. For instance, evolutionary distances are simply the sum of the edge
length; the edge set of Pairwise Compatibility Graphs requires the path
length (i.e., sum of edge-weights) to fall between given bounds
\cite{PCGsurvey}; a pair of genes are orthologs, a key relation in
functional genomics, if their last common ancestor $\lca\lc{T}(x,y)$ is
labeled as speciation; a directed xenology relation is defined by asking
whether there is a ``transfer edge'' on the path between $\lca\lc{T}(x,y)$
and $y$. 

In all these examples the mathematical interest is in the inverse
problem. Given a relation or a set of relations and a rule relating labeled
trees to the relation(s), one asks (i) when does a tree $T$ exist that
explains the given relation, (ii) is there a unique explaining tree $T$
that is minimal in some sense (usually edge contraction), and (iii) can a
(minimal) explaining tree be constructed efficiently from the given
data. For the vertex-labeled case, symbolic ultrametrics \cite{Boecker:98}
and 2-structures \cite{ER2:90,Hellmuth:17a} provide a comprehensive
answer. Edge labels also have been studied extensively.  For distances, the
\emph{4-point condition} \cite{Buneman:71} characterize the ``additive''
metrics deriving from trees. For rare events, where $x\sim y$ if they are
separated by exactly one event, a complete characterization was provided in
\cite{Hellmuth:18b}.  For PCGs (which exclude the possibility of no event
along an edge), on the other hand, only partial results are known
\cite{PCGsurvey}. 

\begin{figure}[tbp]
  \begin{center}
    \includegraphics[width=0.95\textwidth]{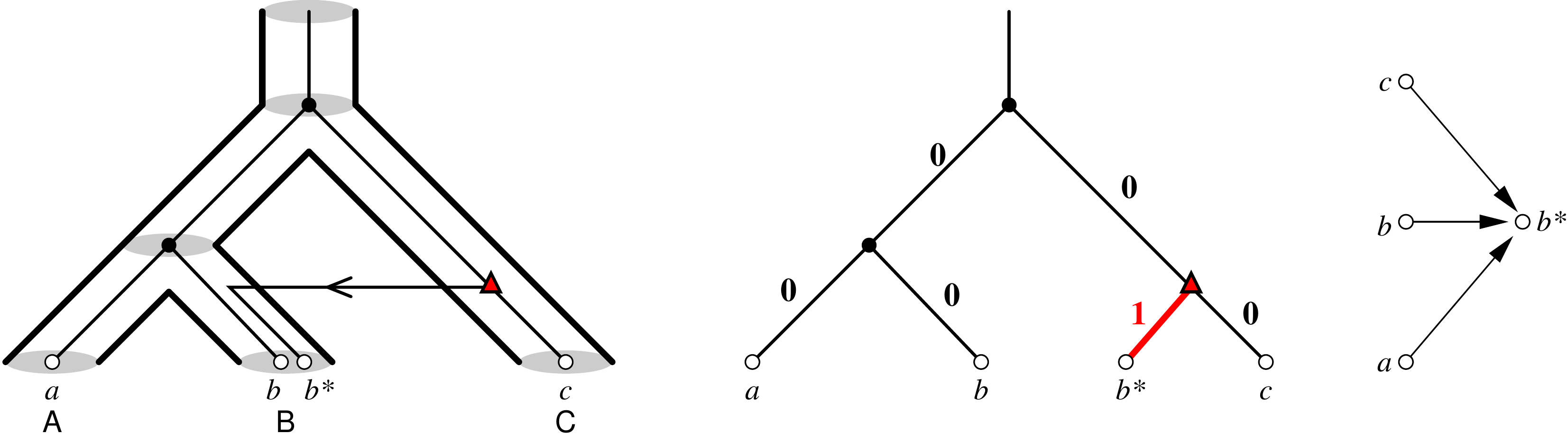}
  \end{center} 
  \caption{The evolution of gene families is modeled as an embedding
      of a gene tree (thin lines, and r.h.s.\ panel, with four genes $a$,
      $b$, $b^*$ and $c$) into a species tree (shown as tubes with fat
      outlines with three species $\mathsf{A}$, $\mathsf{B}$, and
      $\mathsf{C}$). At each speciation (gray ellipses), each gene present
      in the genome is transmitted into both descending lineages,
      corresponding to a speciation even in the gene tree (shown as
      $\bullet$).  Horizontal gene transfer consists in a duplication of a
      gene, one copy of which ``jumps'' into a different lineage. The
      corresponding edge in the gene tree is marked with label $1$.
      The graph representation of the directed Fitch relation
        $\to= \{(a,b^*), (b,b^*), (c,b^*)\}$ is shown on the right hand
        side.  The underlying symmetrized Fitch relation $\sim$ is obtained
        from $\to$ by adding $(x,y)$ to $\sim$ if and only if
        $(x,y)\in \to$ or $(y,x)\in \to$. Figure reproduced from
        \cite{Hellmuth2019gf}.}
  \label{fig:intro}
\end{figure}

In this contribution we are interested in a generalization of Fitch
  relations. These relations were introduced to model so-called horizontal
  gene transfer (HGT) and formalized in \cite{Geiss:18a,Hellmuth:18a}
  based on Walter M.\ Fitch's definition of xenology \cite{Fitch:00}.  Two
  genes $x$ and $y$ are in the directed Fitch relation ($x\to y$) if there
  is at least one event (e.g.\ HGT) on the path between $\lca\lc{T}(x,y)$
  and $y$ \cite{Geiss:18a} and they are in the symmetrized (or undirected)
  Fitch relation ($x\sim y$) if there is at least one event on the path
  between $x$ and $y$ \cite{Hellmuth:19F}. While directed Fitch relations
  ``implicitly'' contain information about the directions of edges in a
  rooted tree, this information is lost in symmetrized Fitch relations.  In
  particular, every symmetrized Fitch relation can be explained by a rooted
  tree if and only if it can be explained by an unrooted tree.
  \cref{fig:intro} illustrates these concepts.  Modeling different types of
  events by different labels yields a collection of (possibly non-disjoint)
  Fitch relations or, equivalently, multi-edge-colored graphs
  \cite{Hellmuth:2019d} that can be used e.g.\ to distinguish genomic
  locations where the horizontally transferred gene copy is inserted, and
  adds to the information that is already provided by a single Fitch
  relation. 

The directed Fitch relations corresponds to a certain subclass of directed
cographs, which are explained by unique least-resolved trees
\cite{Geiss:18a,Hellmuth:19F}. The latter construction was further
generalized to Fitch maps, or, equivalently, sets of Fitch relations, for
every value of the label set. This imposes additional constraints beyond
the obvious fact that one must have a Fitch relation for each label; again
there is a unique least-resolved tree for every Fitch map
\cite{Hellmuth2019gf}. 
Symmetrized Fitch
  relations are of particular interest, as in practice, they can -- to a
  large extent -- be directly inferred from sequence similarities (best
  matches) on genomic data \cite{Schaller:20}.
The symmetrized Fitch relation coincide with the complete multi-partite
graphs \cite{Hellmuth:18a}. This begs the question whether symmetrized
Fitch maps can be understood as simple superpositions of complete
multi-partite graphs.

Our objects of interest -- the Fitch relations and Fitch maps -- are
  named in honor of Walter M.\ Fitch (1929-2011) due to his seminal work on
  xenology. To the best of our knowledge, they bear no relationship to
  Fitch's algorithm \cite{Fitch:71}, which -- invented by the same W.M.\
  Fitch -- solves the ``small parsimony problem'', a combinatorial
  optimization problem concerned with minimizing the the total number
  differences between adjacent vertex labels (character states) in given
  phylogenetic trees.  Since Fitch maps ask for the existence of an edge
  color rather then the number of edges carrying a given color, there is no
  apparent relation to a parsimony criterion for events (i.e., colors) on
  the edges.
  
Symmetric Fitch maps are distantly related to perfect phylogenies. In
  this important model of mathematical phylogenetics, each type of event
  (change of the state of a character) occurs exactly once, see e.g.\
  \cite{Steel:92,Kannan:97,FernandezBaca:01}. A perfect phylogeny is
  illustrated by the insertion of repetitive elements at a certain loci,
  from which they cannot be lost.  In Fitch maps, each color may appear
  multiple times. Different edges of the same color may then be interpreted
  as insertions of members of a given repeat family at different
  loci. Different colors would then correspond to different repeat
  families. In the classical setting of phylogenetics the character states
  (i.e., presence or absence at each locus) are known for all leaves of the
  phylogenetic tree. A symmetric Fitch map, however, withholds the detailed
  information on the number and states of the characters, and instead only
  provides coarse-grained information on differences between sets of
  characters as observables. The recognition problem for Fitch maps
  therefore asks whether there exists a perfect phylogeny with an unknown
  number of binary characters for each color, that ``explains'' the available
  information.

The main result of this contribution states that a collection of binary
relations is a symmetrized Fitch map if and only if each of them is a
complete multi-partite graph and a certain set of subsplits defined by
so-called complementary neighborhoods is compatible.  This characterization
has important consequence on the computational complexity. While
symmetrized Fitch relations (or equivalently, undirected Fitch
  graphs) as well as directed Fitch maps can be recognized in polynomial
time, this is no longer the case for symmetrized Fitch map; we show that
their recognition problem is \NP-complete.  The restriction to maps where
each pair of leaves $(x,y)$ has at most one label, however, remains
polynomial. In particular, this work complements the results established in
\cite{Hellmuth:2019d,Hellmuth2019gf}.

\section{Preliminaries}

\paragraph{Basic Notation}
For a finite set $\X$ we write
$\irr{\X \times \X} \coloneqq \X\times\X \setminus\{(x,x)\colon x\in\X\}$,
and $\binom{\X}{k} \coloneqq \{ \X' \subseteq \X \colon |\X'|=k \}$. 
The set $\PS{\X}$ denotes the \emph{power set} of $\X$.  A
\emph{partition} of $\X$ is a collection of pairwise disjoint non-empty
sets $\X_1,\ldots,\X_k$ with $k\ge 1$ such that
$\X=\X_1\cupdot\ldots\cupdot\X_k$.

We consider undirected graphs $G=(V,E)$ with finite vertex set $V(G)=V$ and
edge set $E(G)=E\subseteq \binom{V}{2}$, i.e., without loops and multiple
edges. The \emph{complete graph} $K_{|V|}$ has vertex set $V$ and edge set
$E=\binom{V}{2}$.
Hence, $K_1$ denotes the single vertex graph and $K_2$
consist of two vertices and the connecting edge.  The vertex degree
$\deg\lc{G}(v)$ of $v\in V$ is the number of its adjacent vertices.  A
graph $H=(W,F)$ is a \emph{subgraph} of $G=(V,E)$, denoted by
$H\subseteq G$, if $W\subseteq V$ and $F\subseteq E$.

A subset $W\subseteq V$ is an \emph{independent set} of $G=(V,E)$, if
  $\{x,y\}\notin E$ for all $x,y\in W$.  $G=(V,E)$ is a \emph{complete
  multi-partite} graph if there is a partition $V_1,\dots, V_k$, $k\geq 1$
of $V$ such that $\{x,y\}\in E$ if and only if $x\in V_i$ and $y\in V_j$
with $i\neq j$. Thus, each part $V_i$ is a maximal independent set. A
graph $G$ is a complete multi-partite graph if and only if it does not
contain $K_1+K_2$, the \emph{disjoint union} of $K_1$ and $K_2$, i.e., the
graph with three vertices and a single edge as an induced subgraph, see
e.g.\ \cite{Zverovich:99}.

\paragraph{Trees}
An \emph{(unrooted) tree} $T=(V,E)$ is a connected, cycle-free graph.  In a
tree, there is a unique path $P\lc{T}(v,w)$ connecting any two vertices
$v,w\in V(T)$. A vertex $v\in V(T)$ with $\deg\lc{T}(v)=1$ is a
\emph{leaf}, otherwise it is an \emph{inner} vertex. The set of
    inner vertices is denoted by $\mathring{V}(T)$. Analogously, an edge
  $e=\{v,w\}\in E(T)$ with $v,w \in \mathring{V}(T)$ is an \emph{inner}
  edge, and an \emph{outer} edge, otherwise. The set of inner edges of
    a tree $T$ is denoted by $\mathring{E}(T)$.  The tree $T$ is
\emph{binary} if $\deg\lc{T}(v)=3$ for every
$v \in V(T)\setminus\mc{L}(T)$.  An (unrooted) tree $T$ is
\emph{phylogenetic} if $\deg\lc{T}(v)\ge 3$ for every vertex
$v \in V(T)\setminus\mc{L}(T)$.  A \emph{star tree} is a tree that has
  exactly one inner vertex and at least two leaves. Moreover, we say a tree
  $T$ is \emph{less resolved than a tree $T'$}, denoted by $T<T'$, if
  $T$ can be obtained from $T'$ by a non-empty sequence of
  edge-contractions.

\begin{remark}
  From here on we consider only phylogenetic trees, and refer to them
  simply as \emph{trees}.
\end{remark}

\paragraph{Subsplits and Quartets}
A \emph{subsplit $A|B$} on a set $\X$ is an unordered pair of two disjoint
and non-empty subsets $A,B \subseteq \X$, i.e.\ $A|B=B|A$.  A subsplit
$A|B$ is \emph{trivial} if $\min\{|A|,|B|\}=1$, and it is a \emph{quartet}
if $|A|=|B|=2$. In the latter case we write $ab|cd$ instead of
$\{a,b\}|\{c,d\}$. A subsplit $A|B$ on $\X$ is a \emph{split} on $\X$ if
$A\cup B=\X$. A subsplit $A|B$ on $\X$ is \emph{displayed} by a tree $T$
with $\mc{L}(T)=\X$ if there is an edge $e \in E(T)$ such that
$A\subseteq\mc{L}(T_1)$ and $B\subseteq\mc{L}(T_2)$, where $T_1$ and $T_2$
are the connected components of
$T\setminus e \coloneqq(V(T),E(T)\setminus \{e\})$. In this case we call
$e$ a \emph{splitting} edge w.r.t.\ $A|B$.  Clearly, removal of an edge in
$T$ yields always a split $\mc{L}(T_1)|\mc{L}(T_2)$ that is displayed by
$T$.  Hence, a subsplit $A|B$ is displayed by $T$ if there is a split
$A'|B'$ in $T$ with $A\subseteq A'$ and $B\subseteq B'$.  A set
$\mathcal{S}$ of subsplits is called \emph{compatible} if there is a tree
$T$ that displays every subsplit in $\mathcal{S}$.  The set
$\mathcal{S}(T)$ comprises all splits on $\X$ displayed by $T$ and the set
$\Q(T)$ comprises all quartets that are displayed by $T$.

The relation between trees and split systems is captured by the following
well-known result \cite{Buneman:71}, see \cite[Section 3.1]{Semple2003} for
a detailed discussion. In the setting of $X$-trees, the taxa $\X$ are
  mapped to vertices of the $T$ with degree at most $2$ by a not
  necessarily injective map $p:X\to V(T)$. Since in our setting there is a
  one-to-one correspondence of $\X$ and the leaves of $T$, i.e., $p$ is
  injective, the split system $\mathcal{S}$ necessarily contains all
  trivial splits $\{x\}|\X\setminus\{x\}$ with $x \in \X$.
\begin{proposition}[Splits-Equivalence Theorem]
  Let $\mathcal{S}$ be a collection of splits on $\X$ that contains
    all trivial splits. Then, there is a tree $T$ with leaf set $\X$ such
  that $\mathcal{S} = \mathcal{S}(T)$ if and only if for all pairs of
  distinct splits $A_1|B_1,A_2|B_2\in\mathcal{S}$ at least one of the four
  intersections $A_1 \cap A_2$, $A_1\cap B_2$, $B_1 \cap A_2$ and
  $B_1 \cap B_2$ is empty.  Moreover, if such a tree exists, then $T$ is
  unique up to isomorphism.
\label{thm:split-equi}
\end{proposition}

For later reference we state a simple consequence of \Cref{thm:split-equi}.
\begin{corollary} \label{cor:subsplit-equi} 
  Let $\mathcal{S}$ be a collection of subsplits on $\X$. If there are two
  subsplits $A_1|A_2$ and $B_1|B_2$ in $\mathcal{S}$ such that all four
  intersections $A_1 \cap B_1$, $A_1\cap B_2$, $A_2 \cap B_1$ and $A_2 \cap
  B_2$ are non-empty, then $\mathcal{S}$ is not compatible.
\end{corollary} 
\begin{proof}
  Let $\mc{S}$ be a collection of subsplits on $\X$, and suppose that are
  two subsplits $A_1|B_1$ and $A_2|B_2$ in $\mathcal{S}$ such that none of
  the sets $A_1 \cap A_2$, $A_1\cap B_2$, $B_1 \cap A_2$ and $B_1 \cap B_2$
  is empty. Assume for contradiction that $\mc{S}$ is compatible, i.e.,
  there is a tree $T$ that displays $S$. Thus, there is a split $A'_1|A'_2$
  and a split $B'_1|B'_2$ in $T$ such that $A_1\subseteq A'_1$,
  $A_2\subseteq A'_2$, $B_1\subseteq B'_1$ and $B_2\subseteq B'_2$.
  However, by assumption all four intersections
  $A_i'\cap B_j'\supseteq A_i\cap B_j\ne\emptyset$ with $i,j \in \{1,2\}$,
  and hence, by \Cref{thm:split-equi}, such a tree $T$ cannot exist.
  Therefore, $\mc{S}$ is not compatible.
\end{proof}

\section{Symmetrized Fitch maps}

\begin{definition}
  Let $M$ be an arbitrary finite set of colors. An \emph{edge-labeled tree
    $(T,\lambda)$ on $\X$ (with $M$)} is a tree $T=(V,E)$ with
  $\mc{L}(T)=\X$ together with a map $\lambda: E \to \PS{M}$.
\end{definition}
We will often refer to the map $\lambda$ as the \emph{edge-labeling} and
call $e$ an \emph{$m$-edge} if $m \in\lambda(e)$, and an $\emptyset$-edge
if $\lambda(e)=\emptyset$.  Note that the choice of $m\in \lambda(e)$ may
not be unique and an edge can be both, an $m$- and an $m'$-edge at the same
time.

\begin{definition}
  \label{d:fitch-map}
  A map $\eXM$, where $\X$ is a non-empty set of ``leaves'' and $M$ is a
  non-empty set of ``colors'', is a \emph{symmetrized Fitch map} if there
  is an edge-labeled tree $(T,\lambda)$ with leaf set $\X$ and edge
  labeling $\lambda: E(T)\to \PS{M}$ such that
  for every pair $(x,y)\in \irr{\X\times \X}$ it holds that
  \begin{equation*}
    m \in \eps(x,y) \iff \textnormal{ there is an }
    m\textnormal{-edge on the path from } x \textnormal{ to } y.
  \end{equation*}
  In this case we say that $\eXM$ \emph{explains} $(T,\lambda)$.
\end{definition}
Every symmetrized Fitch map is symmetric, i.e., $\eps(x,y)=\eps(y,x)$ for
every distinct $x,y \in \X$.  Furthermore, every symmetric map $\eXM$ with
$|\X|=2$ is a symmetrized Fitch map.

\begin{remark}
  From here on we assume w.l.o.g.\ that $\eps$ is symmetric and
  $|\X|\geq 3$.
\end{remark}

\begin{figure}[t]
  \begin{center}
    \includegraphics[width=0.7\textwidth]{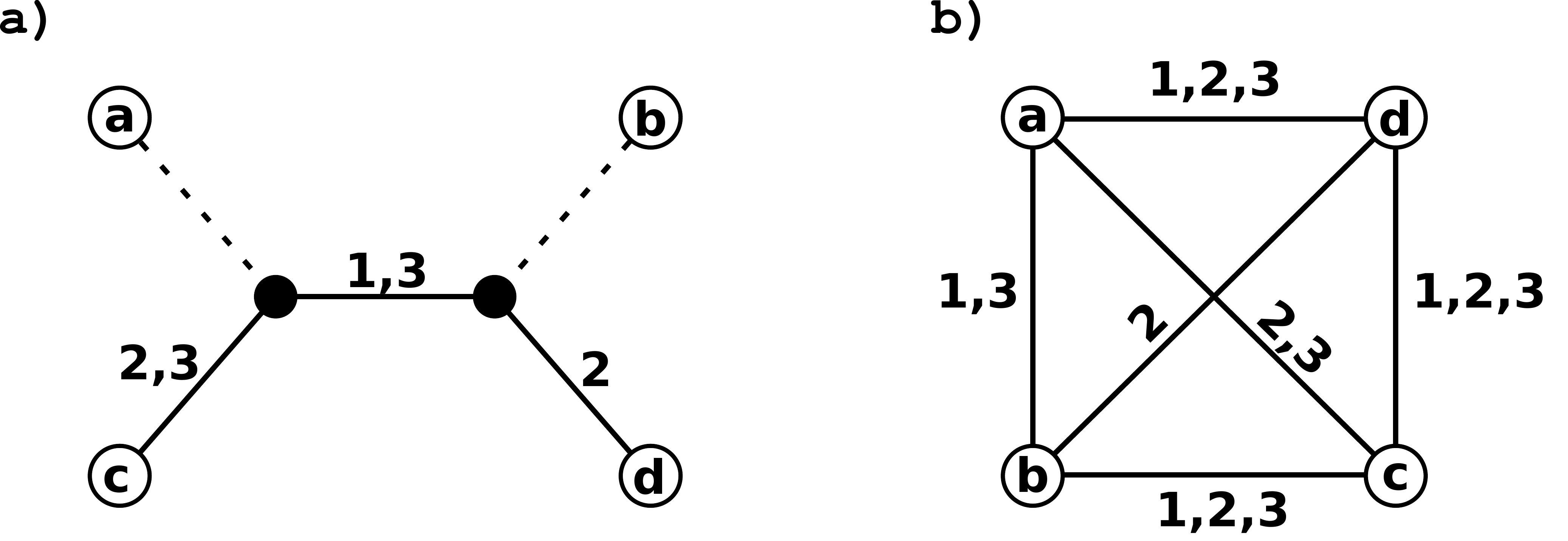}
  \end{center}
  \caption{The edge-labeled tree $(T,\lambda)$ with leaf set
    $\mc{L}(T)=\{a,b,c,d\}=: \X$ on the left-hand side explains the
    symmetrized Fitch map $\eXM$ with the color set $M=\{1,2,3\}$ on the
    right-hand side.  Dashed-lined edges $e$ in $T$ have label
    $\lambda(e)=\emptyset$. Moreover, an edge $\{x,y\}$ in the the
    right-hand side graph has label $i$ if and only if $i\in \eps(x,y)$.
    By definition, symmetrized Fitch maps are symmetric, i.e.,
    $m \in \eps(x,y)$ if and only if $m \in \eps(y,x)$ for all $m\in M$.
    However, symmetrized Fitch maps are not transitive in general. To see
    this, observe that $1 \in \eps(a,b)$ and $1 \in \eps(b,c)$ but
    $1 \notin \eps(a,c)$.}
  \label{fig:exmpl}
\end{figure}

\begin{figure}[t]
  \begin{center}
    \includegraphics[width=0.8\textwidth]{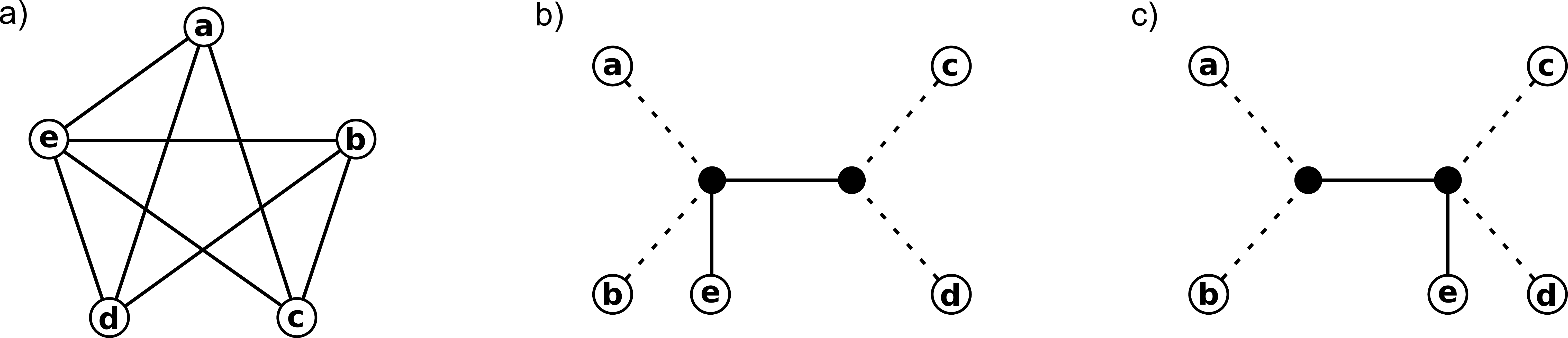}
  \end{center}
  \caption{Let $\eXM$ be a symmetric map with
    $\X\coloneqq \{a,b,c,d,e\}$ and $M\coloneqq \{m\}$, where for every
    distinct $x,y \in \X$ we have $m \in \eps(x,y)$ if and only if the
    shown graph in Panel a) contains the edge $\{x,y\}$.  Moreover, there
    are two edge-labeled trees shown in Panel b) and c), where solid lines
    and dashed lines represent the edge-label $\{m\}$ and $\emptyset$,
    respectively.  We observe that both edge-labeled trees explains
    $\eps$. Thus, $\eps$ is a (monochromatic) symmetrized Fitch relation.
    For instance, $m \notin \eps(a,b)\cup \eps(c,d)$ but
      $m \in \eps(a,c)$ imply that every edge-labeled tree, which explains
      $\eps$, needs at least one inner edge.  Thus, these two trees have
    the fewest numbers of vertices among all trees that may explain $\eps$
    and are known as so-called ``minimally-resolved'' trees.  The latter
    arguments imply that minimally-resolved trees need not to be unique; a
    fact that has also been observed in \cite{Hellmuth:18a}.}
  \label{fig:least-resolved}
\end{figure}

\Cref{fig:exmpl} provides an illustrative example of a symmetrized Fitch
map $\eXM$ and one of its corresponding edge-labeled trees
$(T,\lambda)$.  In particular, \Cref{fig:least-resolved} shows
  that the corresponding edge-labeled trees for $\eps$ may not be unique in
  general.
Every map $\eXM$ can also be interpreted as a set of $|M|$ not
necessarily disjoint binary relations (or equivalently graphs) on $\X$
defined by the sets $\{(x,y) \in\irr{\X\times \X} \colon m \in \eps(x,y)\}$
of pairs (or equivalently undirected edges) for every fixed color $m\in M$.
\begin{definition}
  The \emph{graph-representation} of a map $\eXM$ w.r.t.\ a color
  $m \in M$ is the (undirected) graph $\G\lc{m}(\eps)$ with the vertex set
  $V(\G\lc{m}(\eps))\coloneqq \X$ and the edge set
  $E(\G\lc{m}(\eps))\coloneqq \left\{ \{x,y\} \in {\X \choose 2} \colon m
    \in \eps(x,y) \right\}.$
\end{definition}
Following the approach by \citet{Hellmuth2019gf}, we start by considering
neighborhoods in this graph representation.
\begin{definition}[{\cite[Def.\ 3.3]{Hellmuth2019gf}}] \label{def:neigh}
  The \emph{(complementary) neighborhood} of vertex $y \in \X$ and a
  given color $m \in M$ w.r.t.\ $\eXM$ is the set
  \begin{equation*}
    \NnotCol{m}[y] \coloneqq \{ x \in \X \setminus\{y\} : m \notin
    \eps(x,y) \} \cup \{y\}
  \end{equation*}
\end{definition}
We write $\Ns{m}[\eps]\coloneqq \{ \NnotCol{m}[y]\colon y\in \X \}$ for the
set of complementary neighborhoods of $\eps$ and a particular color
$m \in M$.  Note that there might be distinct leaves $y,y'\in \X$ or
distinct colors $m,m' \in M$ such that $\NnotCol{m}[y]=\NnotCol{m'}[y']$.
Moreover, we emphasize that $\Ns{m}[\eps]$ is not a multi-set, i.e.\ if
$\NnotCol{m}[y]=\NnotCol{m'}[y']$ but $y\neq y'$ or $m\neq m'$, then they
only contribute once to $\Ns{m}[\eps]$.

\subsection{Characterization of monochromatic symmetrized Fitch maps}

A map $\eXM$ is \emph{monochromatic} if $\eps(x,y) = \{m\}$ or
$\eps(x,y) = \emptyset$ for all distinct $x,y\in \X$ and some fixed
color $m\in M$. Hence, for monochromatic maps we can assume w.l.o.g.\ that
$|M|=1$. Monochromatic symmetrized Fitch maps are equivalent to the
``undirected Fitch graphs'' studied by \citet{Hellmuth:18a}.

For later reference we briefly recall some key results for this special
case.
\begin{lemma}[{\cite[Lemma 0.3 \& Thm.\ 0.5]{Hellmuth:18a}}]
  \label{lem:mono-k1+k2}
  Let $\eXM$ be a monochromatic map with $M=\{m\}$.  Then, the following
  statements are equivalent:
  \begin{enumerate}%[noitemsep]
  \item $\eps$ is a (monochromatic) symmetrized Fitch map.
  \item $\G\lc{m}(\eps)$ does not contain a $K_1+K_2$ as an induced
    subgraph.
  \item $\G\lc{m}(\eps)$ is a complete multi-partite graph.
  \end{enumerate}
\end{lemma}
Using \Cref{lem:mono-k1+k2}, we can derive the following alternative
characterization:
\begin{proposition}\label{prop:mono-partition}
  Let $\eXM$ be a monochromatic map with $M=\{m\}$.  Then, the following
  statements are equivalent:
  \begin{enumerate}%[noitemsep]
  \item $\eps$ is a (monochromatic) symmetrized Fitch map.
  \item For every three pairwise distinct $a,b,c \in \X$ with
    $m \notin \eps(a,b)$ and $m \notin \eps(b,c)$, we have
    $m \notin \eps(a,c)$.
  \item $\Ns{m}[\eps]$ is a partition of $\X$.
  \end{enumerate}
\end{proposition}
\begin{proof}
  Let $\eXM$ be a monochromatic map with $M=\{m\}$. In the following will
  make frequent use of the fact that $\eps(a,b)=\eps(b,a)$ and, therefore,
  $m\in \eps(a,b)$ if and only if $\{a,b\}\in E(\G\lc{m}(\eps))$.

  First, assume that Statement~(1) is satisfied.  \Cref{lem:mono-k1+k2}
  implies that $\G\lc{m}(\eps)$ does not contain a $K_1+K_2$ as an induced
  subgraph.  Hence, for arbitrary pairwise distinct $a,b,c \in \X$ with
  $m \notin \eps(a,b)$ and $m \notin \eps(b,c)$, it must hold that
  $m \notin \eps(a,c)$. Thus, Statement~(2) holds.
	
  Now, assume that Statement~(2) is satisfied.  Recall that the set
  $\Ns{m}[\eps]$ is a partition of $\X$ if $\Ns{m}[\eps]$ is a collection
  of pairwise disjoint non-empty sets $N_1,\ldots,N_k$ such that
  $\X=N_1\cupdot\ldots\cupdot N_k$.  Since $y \in \NnotCol{m}[y]$, we
  conclude that every neighborhood in $\Ns{m}[\eps]$ is non-empty and that
  $\bigcup_{y\in \X} \NnotCol{m}[y]=\X$.  To this end, let $y,y'\in X$ be
  two distinct vertices that satisfy
  $\NnotCol{m}[y]\cap\NnotCol{m}[y']\ne \emptyset$.  Thus, we must verify
  that $\NnotCol{m}[y] = \NnotCol{m}[y']$.  Moreover, we can assume
  w.l.o.g.\ that $|\NnotCol{m}[y]|\le |\NnotCol{m}[y']|$.  Now, we continue
  to show that $m \notin \eps(y,y')=\eps(y',y)$. To this end, we assume for
  contradiction that $m \in \eps(y,y')=\eps(y',y)$. Therefore,
  $y \notin \NnotCol{m}[y']$ and $y' \notin \NnotCol{m}[y]$. Thus,
  $y,y' \notin \NnotCol{m}[y]\cap \NnotCol{m}[y']$. This, together with
  $\NnotCol{m}[y]\cap\NnotCol{m}[y']\ne \emptyset$, implies that there is a
  vertex $x \in \NnotCol{m}[y]\cap\NnotCol{m}[y']$ such that $x,y$ and $y'$
  are pairwise distinct. However, $m \notin \eps(x,y)=\eps(y,x)$ and
  $m \notin \eps(x,y')$. In summary, we have $m \notin \eps(y,x)$,
  $m \notin \eps(x,y')$ and $m \in \eps(y,y')$; a contradiction to
  Statement~(2). Thus, $m \notin \eps(y,y')=\eps(y',y)$. The latter implies
  that $\{y,y'\}\subseteq\NnotCol{m}[y]$.  Now, let
  $x \in \NnotCol{m}[y']$.  If $x \in \{y,y'\}$, then we have
  $x \in \{y,y'\}\subseteq\NnotCol{m}[y]$.  Moreover, if
  $x \notin \{y,y'\}$, then $x,y$ and $y'$ are pairwise distinct.  In this
  case, $m \notin \eps(x,y')$ and $m \notin \eps(y',y)$ together with
  Statement (2) implies that $m \notin\eps(x,y)$. Therefore,
  $x \in \NnotCol{m}[y]$.  In either case, we have $x \in
  \NnotCol{m}[y]$. Thus, $\NnotCol{m}[y']\subseteq\NnotCol{m}[y]$.  This,
  together with $|\NnotCol{m}[y]|\le |\NnotCol{m}[y']|$, implies that
  $\NnotCol{m}[y]=\NnotCol{m}[y']$.  Therefore, Statement~(3) is true.
	
  Finally, we show that Statement~(3) implies Statement~(1).  Using
  contraposition, we assume that $\eps$ is not a symmetrized Fitch map.
  Then, we conclude by \Cref{lem:mono-k1+k2} that $\G\lc{m}(\eps)$ contains
  an $K_1+K_2$ as an induced subgraph.  Let $\G\lc{m}(\eps)[\{a,b,c\}]$ be
  an induced subgraph that is isomorphic to $K_1+K_2$.  We can assume
  w.l.o.g.\ that $m \notin \eps(a,b)$, $m \notin \eps(a,c)$ and
  $m \in \eps(b,c)$.  The latter implies that $b \notin \NnotCol{m}[c]$.
  This, together with $b \in \NnotCol{m}[b]$, implies that
  $\NnotCol{m}[b]\ne \NnotCol{m}[c]$.  Moreover, we have
  $a \in \NnotCol{m}[b]\cap \NnotCol{m}[c]$.  Taken the latter arguments
  together, we observe that $\Ns{m}[\eps]$ cannot be a partition of $\X$.
  Thus, if Statement~(3) is satisfied, then Statement~(1) must be satisfied
  as well.
\end{proof}

A natural special case is to consider maps $\eXM$ that assign to each pair
$(x,y)$ at most one label. In this case, $\eps$ reduces to a map
$\eps \colon \irr{\X \times \X} \to M\cup \{\emptyset\}$.
\begin{proposition} \label{prop:restricted-mono} The map $\eXM$ is a
  symmetrized Fitch map that satisfies $|\eps(x,y)|\le 1$ for all distinct
  $x,y\in X$ if and only if $\eps$ is a monochromatic symmetrized Fitch
  map.
\end{proposition}
\begin{proof}
  Clearly, every monochromatic symmetrized Fitch map $\eps$ is a
  symmetrized Fitch map with $|\eps(x,y)|\le 1$ for all distinct
  $x,y\in X$. Now, suppose that $\eXM$ is a symmetrized Fitch map that
  satisfies $|\eps(x,y)|\le 1$.  Then, assume for contradiction that $\eps$
  is not monochromatic. Thus, there are leaves $a,b,c,d\in \X$ with
  $\eps(a,b)=\{m\}$ and $\eps(c,d)=\{m'\}$ for distinct $m,m'\in M$.  Since
  $\eps$ is a symmetrized Fitch map, there is an edge-labeled tree
  $(T, \lambda)$ that explains $\eps$.  The latter two arguments imply that
  $T$ contains an $m$-edge $e$ and $m'$-edge $f$.  Now, consider a
  vertex-maximal path $P$ in $T$ that contains $e$ and $f$.  Clearly, $P$
  must contain two leaves $x,y\in \X$ as its end-vertices.  But then
  $m,m'\in \eps(x,y)$ implies $|\eps(x,y)|> 1$; a contradiction.
\end{proof}

\begin{figure}[t]
  \begin{center}
    \includegraphics[width=\textwidth]{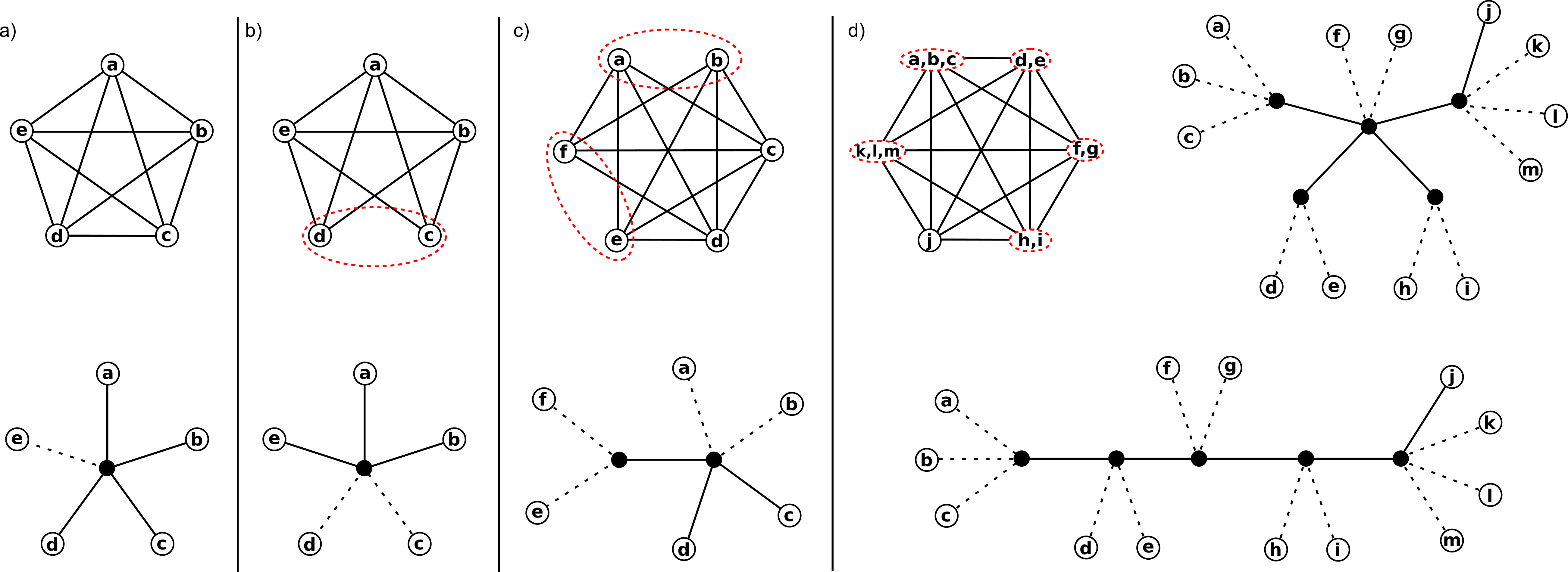}
  \end{center}
  \caption{Graph representations $\G\lc{m}(\eps)$ (Fitch graphs) of
      several monochromatic symmetrized Fitch maps $\eXM$, $M=\{m\}$
      (\emph{upper parts}) together with edge-labeled trees $(T,\lambda)$
      that explain $\eps$ (drawn next to the respective $\G\lc{m}(\eps)$).
      The Fitch graphs are complete bipartite graphs and thus, are
      identified by the set $\mc{I}$ of their maximal independent sets.
      All maximal independent sets of size at least two are highlighted by
      red dashed-lined ellipses.
      All trees satisfy the conditions in
      \cref{def:minimally-resolved-tree} and are thus, by
      \cref{prop:charact-least-resolvedness} least-resolved and, by
      \cref{cor:min-res}, minimally-resolved for each $\eps$.
      }
\label{fig:diam4}
\end{figure}

Finally, we characterize least-resolved trees for a monochromatic
  symmetrized Fitch map $\eps$: An edge-labeled tree $(T^*,\lambda^*)$ is
  \emph{least-resolved} for $\eps$ if there is no tree $T<T^*$ and no
  labeling $\lambda$ such that $(T,\lambda)$ also explains $\eps$.  Such
  trees can be constructed using the fact that $\G\lc{m}(\eps)$ is
  identified by the set $\mc{I}$ of its maximal independent sets since it
  is a complete multi-partite graph (cf.\ \cref{lem:mono-k1+k2}). As
  remarked above, we assume $|X|\geq 3$ to avoid trivial cases.  
\begin{definition}\label{def:minimally-resolved-tree}
  Let $\eXM$ be a monochromatic symmetrized Fitch map with $M=\{m\}$ and
  $|X|\geq 3$ with graph-representation $\G\lc{m}(\eps)$. Then,
  $\mathcal{T}_{\eps}$ is the collection of all edge-labeled trees
  $(T,\lambda)$ on $\X$ that satisfy the following properties:
  \begin{enumerate}
  \item  If $\mc{I}_2 \coloneqq \{I\in \mc{I}: |I|\geq 2\} =\emptyset$,
    i.e.\ if $\G\lc{m}(\eps) \simeq K_{|\X|}$, then $(T,\lambda)$ is the
    star tree on $\X$, and all edges $e \in E(T)$ with one possible
    exception are $m$-edges.

  \item If $\mc{I}_2\ne\emptyset$, then 
      \begin{enumerate}
    \item $|\mathring{V}(T)|=|\mc{I}_2|$;
    \item for each $I \in \mc{I}_2$ there is a unique inner vertex
      $v\lc{I} \in \mathring{V}(T)$ such that $\{x,v\lc{I}\} \in E(T)$ for
      every $x \in I$;
    \item every outer edge $\{x,v\} \in E(T)$ with
      $x \in \bigcup_{I\in \mc{I}_2}I$ is labeled by
      $\lambda(\{x,v\})=\emptyset$;
    \item every outer edge $\{x,v\} \in E(T)$ with
      $x\in \X\setminus \bigcup_{I\in \mc{I}_2} I$ is labeled by
      $\lambda(\{x,v\})=\{m\}$;
    \item every inner edge $e \in \mathring{E}(T)$ is labeled by
    $\lambda(e)=\{m\}$.
  \end{enumerate}
\end{enumerate}
\end{definition}
Note that there are no restrictions on the arrangement of the inner edges
in \cref{def:minimally-resolved-tree} as long as $(T,\lambda)$ results in a
phylogenetic tree. For instance, the inner edges could be arranged as a
star-graph or as a path. See \cref{fig:diam4} for an illustrative
example.

\begin{proposition} \label{prop:charact-least-resolvedness} Let $\eXM$ be a
  monochromatic symmetrized Fitch map with $M=\{m\}$, and let
  $\mathcal{T}_{\eps}$ be the collection of trees as specified in
  \cref{def:minimally-resolved-tree}.  Then, $\mathcal{T}_{\eps}$ is the
  set of all least-resolved trees for $\eps$.
\end{proposition}
\begin{proof}
  Let $\eXM$ be a monochromatic symmetrized Fitch map, and
  $\mc{I}, \mc{I}_2$ and $\mathcal{T}_{\eps}$ as specified in
  \cref{def:minimally-resolved-tree}.  First, assume that
  $\mc{I}_2 = \emptyset$, i.e.\ $\G\lc{m}(\eps)\simeq K_{|X|}$.  It is an
  easy exercise to verify that all least-resolved trees for $\eps$ must be
  a star tree, and all edges with one possible exception are $m$-edges.
  Hence, all such trees are, by \cref{def:minimally-resolved-tree}~(1),
  contained in $T_\eps$.
  
  Let us now assume that $\mc{I}_2\neq \emptyset$ and let $(T^*,\lambda^*)$
  be a least-resolved tree for $\eps$. We show that
  $(T^*,\lambda^*)\in T_\eps$.  To this end, we assume first, for
  contradiction, that there is an inner edge $e \in \mathring{E}(T^*)$ with
  $\lambda(e)=\emptyset$, and consider the edge-labeled tree
  $(T',\lambda')$ obtained from $T^*$ by contraction of $e$ and keeping the
  remaining edge labels of $\lambda^*$. Note that $T'$ is still a
  phylogenetic tree. Let $x,y \in \X$ be two distinct leaves.  If
  $\eps(x,y) = \emptyset$, then the path $P\lc{T^*}(x,y)$ can have only
  $\emptyset$-edges. It is easy to see that this property is preserved by
  $(T',\lambda')$. If $\eps(x,y) = \{m\}$, then the path $P\lc{T^*}(x,y)$
  contains an $m$-edge. Since we only contracted the single inner
  $\emptyset$-edge to obtain $T'$, the path $P\lc{T'}(x,y)$ still contains
  this $m$-edge. Consequently, $(T',\lambda')$ explains $\eps$ but
  $T'<T^*$; contradicting the fact that $(T^*,\lambda^*)$ is least-resolved
  for $\eps$. Hence, every inner edge $e\in \mathring{E}(T^*)$ is an
  $m$-edge, and thus, Statement~(2e) holds.  
  
  To see that Statement~(2c) is satisfied, observe that for every
  $x \in \bigcup_{I\in \mc{I}_2}I$ there is an $I \in \mc{I}_2$ with
  $x \in I$, and there is a $y \in I\setminus\{x\}$.  Hence,
  $\eps(x,y)=\emptyset$, and thus, the path $P\lc{T^*}(x,y)$ contains only
  edges $e$ with $\lambda^*(e)=\emptyset$.  In particular, for the outer
  edge $\{x,v\}\in E(T^*)$, we therefore have
  $\lambda^*(\{x,v\})=\emptyset$.  Hence, Statement~(2c) holds.
  
  To see that Statement~(2b) is satisfied, we assume first, for
  contradiction, that there are two vertices $x,y\in I\in \mc{I}_2$ such
  that $\{x,v\lc{I}\}\in E(T^*)$ and $\{y,v'\lc{I}\}\in E(T^*)$ but
  $v\lc{I} \neq v'\lc{I}$.  By Statement~(2e), there is an inner $m$-edge
  contained in $P\lc{T^*}(x,y)$ since $v\lc{I}$ and $v'\lc{I}$ are distinct
  inner vertices.  Thus, $\eps(x,y)=\{m\}\neq \emptyset$; a contradiction
  to $x,y\in I$.  By similar arguments, $\{x,v\},\{x',v\}\in E(T^*)$ with
  $x \in I \in \mc{I}_2$ and $x' \in I' \in \mc{I}_2$ imply
  $x,y \in I\cap I'$, and since $\mc{I}_2$ forms a partition, $I=I'$.
  Thus, the required uniqueness in Statement~(2b) holds.  The previous
  arguments together, imply that Statement~(2b) holds. 
  
  We continue by showing that every inner vertex $v \in \mathring{V}(T^*)$
  is incident to an outer $\emptyset$-edge $\{v,x\}\in E(T^*)$.  If $T$ is
  a star graph, i.e.\ $v \in \mathring{V}(T^*)$ is the only inner vertex,
  then there must be an outer edge $\{v,x\}\in E(T^*)$ with
  $\lambda(\{v,x\})=\emptyset$, since $\mc{I}_2\ne \emptyset$.  Now, assume
  that $T$ has inner edges.  Moreover, assume for contradiction that there
  is an inner vertex $v \in \mathring{V}(T^*)$ such that every outer edge
  $\{v,x\} \in E(T^*)$ has label $\lambda^*(\{x,v\})=\{m\}$.  Since
  $v \in \mathring{V}(T^*)$ and $T$ has inner edges, we can apply
  Statement~(2e) to conclude that there is an inner $m$-edge
  $\{v,w\}\in \mathring{E}(T^*)$.  Now, consider the edge-labeled tree
  $(T',\lambda')$ obtained from $T^*$ by contraction of $\{v,w\}$ and
  keeping the remaining edge labels of $\lambda^*$. Let $\eps'$ be the
  symmetrized Fitch relation explained by $(T',\lambda')$ and $x,y \in \X$
  be chosen arbitrarily.  If $P\lc{T^*}(x,y)$ does not contain the $m$-edge
  $\{v,w\}$, then $P\lc{T'}(x,y)=P\lc{T^*}(x,y)$ and the edge-labels along
  this path remain unchanged. Therefore, $\eps'(x,y) = \eps(x,y)$.  Now,
  suppose that $P\lc{T^*}(x,y)$ contains the edge $\{v,w\}$, and thus,
  $\eps(x,y) = \{m\}$.  If $P\lc{T'}(x,y)$ does not contain any inner edge,
  then either $x$ or $y$ must be incident to $v$ in $T^*$, say
  $\{x,v\}\in E(T^*)$ and thus,
  $\lambda'(\{x,v\}) = \lambda^*(\{x,v\})=\{m\}$.  Hence, $P\lc{T'}(x,y)$
  still contains an $m$-edge, and therefore,
  $\eps'(x,y) = \eps(x,y) = \{m\}$. Otherwise, if $P\lc{T'}(x,y)$ contains
  an inner edge, then it contains in particular an $m$-edge (cf.\
  Statement~(2e)).  Again, $\eps'(x,y) = \eps(x,y) = \{m\}$. Hence, we have
  shown that the symmetrized Fitch relation $\eps'$ that is explained by
  $(T',\lambda')$ and $\eps$ are identical. Consequently, $(T^*,\lambda^*)$
  is not a least-resolved tree for $\eps$; a contradiction. Therefore,
  every inner vertex $v \in \mathring{V}(T^*)$ is incident to an outer
  $\emptyset$-edge $\{v,x\}\in E(T^*)$.
  
  Now, let $\{x,v\}\in E(T^*)$ be an outer edge with
  $x \in \X\setminus\bigcup_{I\in \mc{I}_2}I$.  In particular,
  $x \in \X\setminus\bigcup_{I\in \mc{I}_2}I$ implies that
  $\eps(x,y)=\{m\}$ for every $y \in \X \setminus\{x\}$.  Assume, for
  contradiction, that $\lambda^*(\{x,v\})=\emptyset$.  Due to the choice of
  $x$, every possible outer edge $\{y,v\}\in E(T^*)$ with $x\ne y$ must
  have label $\lambda^*(\{y,v\})=\{m\}$. Let us keep all edge-labels in
  $T^*$ except for $\{x,v\}$, which is relabeled to an $m$-edge. This,
  results in an edge-labeled tree $(T^*,\lambda')$ where $v$ is incident to
  $m$-edges only that still explains $\eps$. But then, $(T^*,\lambda')$ and
  thus, $(T^*,\lambda^*)$ cannot be least-resolved, since for every inner
  vertex $v \in \mathring{V}(T^*)$ there must be an outer $\emptyset$-edge.
  Hence, Statement~(2d) holds.

  It remains to show that Statement (2a) is satisfied. Statement~(2b)
  implies $|\mathring{V}(T)|\ge |\mc{I}_2|$. Assume, for contradiction,
  that $|\mathring{V}(T)|> |\mc{I}_2|$.  Statement~(2b) and
  $|\mathring{V}(T)|> |\mc{I}_2|$ implies that there is an inner vertex
  $v \in \mathring{V}(T^*)$ that is not adjacent to a leaf
  $x \in I\in \mc{I}_2$. Hence, for every outer edge $\{v,x\}\in E(T^*)$ we
  have $x \in \X\setminus\bigcup_{I\in \mc{I}_2}I$, and by Statement~(2d),
  $\{v,x\}$ is an $m$-edge. Hence, $v \in \mathring{V}(T^*)$ is not
  incident to an outer $\emptyset$-edge $\{v,x\}\in E(T^*)$; a
  contradiction. Therefore, $|\mathring{V}(T)|=|\mc{I}_2|$, i.e., Statement
  (2a) is satisfied.

  In summary, every least-resolved $(T^*,\lambda^*)$ satisfies either
  Statement (1) or (2a-e) and thus, $(T^*,\lambda^*) \in \mc{T}_\eps$. In
  particular, all least-resolved trees must have the same number of inner
  vertices, that is, one inner vertex (in case of Statement (1)) or
  $|\mc{I}_2|$ inner vertices (in case Statement (2)). Consequently, all
  least-resolved trees have the same number of vertices, that is, $|\X|+1$
  (Statement (1)) or $|\X|+|\mc{I}_2|$ (Statement (2)).
  
  It remains to show that every tree in $\mathcal{T}_{\eps}$ is also
  least-resolved for $\eps$.  Hence, let
  $(T,\lambda) \in \mathcal{T}_{\eps}$ be an edge-labeled tree. We show
  first that $(T,\lambda)$ explains $\eps$.  To this end, let $x,y\in X$ be
  distinct.  If $\eps(x,y) = \emptyset$, then $x,y$ are contained in the
  same independent set $I$ and, by construction (2b), the path
  $P\lc{T}(x,y)$ contains only the two outer $\emptyset$-edges $\{v_I,x\}$
  and $\{v_I,y\}$.  If $\eps(x,y) = \{m\}$, then $x$ and $y$ are contained
  in distinct independent sets $I,I'\in \mc{I}$, say $x\in I$ and
  $y\in I'$. If $|I|=1$ then, by construction (2c), there is an $m$-edge
  $\{v,x\}$ for some $v\in \mathring{V}(T)$. Similarly, if $|I'|=1$, there
  is an $m$-edge $\{v,y\}$ . If $|I|>1$ and $|I'|>1$, then Statement~(2b)
  implies that $\{x,v\},\{y,v'\}\in E(T)$ with $v\ne v'$.  Hence, the path
  $P\lc{T}(x,y)$ contains an inner edge $e \in \mathring{E}(T)$, which has
  by Statement~(2e) the label $\lambda(e)=\{m\}$.  Hence, in all cases for
  $\eps(x,y) = \{m\}$, the path $P\lc{T}(x,y)$ contains an $m$-edge. In
  summary, $(T,\lambda)$ explains $\eps$.

  It remains to show that $(T,\lambda)$ is least-resolved for $\eps$.
    If $(T,\lambda)$ is not least-resolved for $\eps$ then there is a
    least-resolved tree $(T',\lambda')$ for $\eps$ such that $T'<T$ and
    therefore $|V(T')|<|V(T)|$. As argued above, if $(T^*,\lambda^*)$ is an
    arbitrary least-resolved tree for $\eps$ we have by construction
    $|V(T^*)| = |V(T)|$. Therefore $|V(T')|<|V(T^*)|$, contradicting the
  fact that all least-resolved trees for $\eps$ must have the same number
  of vertices. Consequently, $(T,\lambda)$ is least-resolved for $\eps$.
\end{proof}

In particular, \cref{prop:charact-least-resolvedness} implies the following
\begin{corollary}\label{cor:min-res}
  Let $\eps$ be a monochromatic symmetrized Fitch map.  Then, the
  least-resolved trees for $\eps$ have the same number of vertices and,
  thus, in particular, the minimum number of vertices among all trees that
  explain $\eps$, i.e., they are minimally-resolved trees for $\eps$.
\end{corollary}

\begin{corollary}\label{cor:diam4}
  Every monochromatic symmetrized Fitch map can be explained by an edge
  labeled tree $(T,\lambda)$ of diameter $\diam(T)\le 4$, i.e., the length
  of each path in $T$ is four or less.
\end{corollary}
\begin{proof}
  If $|\mc{I}_2|\leq 1$ and $(T,\lambda)\in \mc{T}_{\eps}$, then, by
  construction, $\diam(T)=2$. Otherwise, if $|\mc{I}_2|> 1$, then there is
  a tree $(T,\lambda)\in \mc{T}_{\eps}$ such that all inner edges of $T$
  share a common vertex as illustrated in Fig. \ref{fig:diam4}. In
  this case, $\diam(T)=3$ if $|\mc{I}_2|=2$ and $\diam(T)=4$ if
  $|\mc{I}_2|>2$.
\end{proof}
\Cref{cor:diam4} can also be obtained from the explicit construction of
rooted trees that explain undirected Fitch graphs \cite{Hellmuth:18a}.

\subsection{Characterization of symmetrized Fitch maps}

\begin{figure}[t]
  \begin{center}
    \includegraphics[width=.2\textwidth]{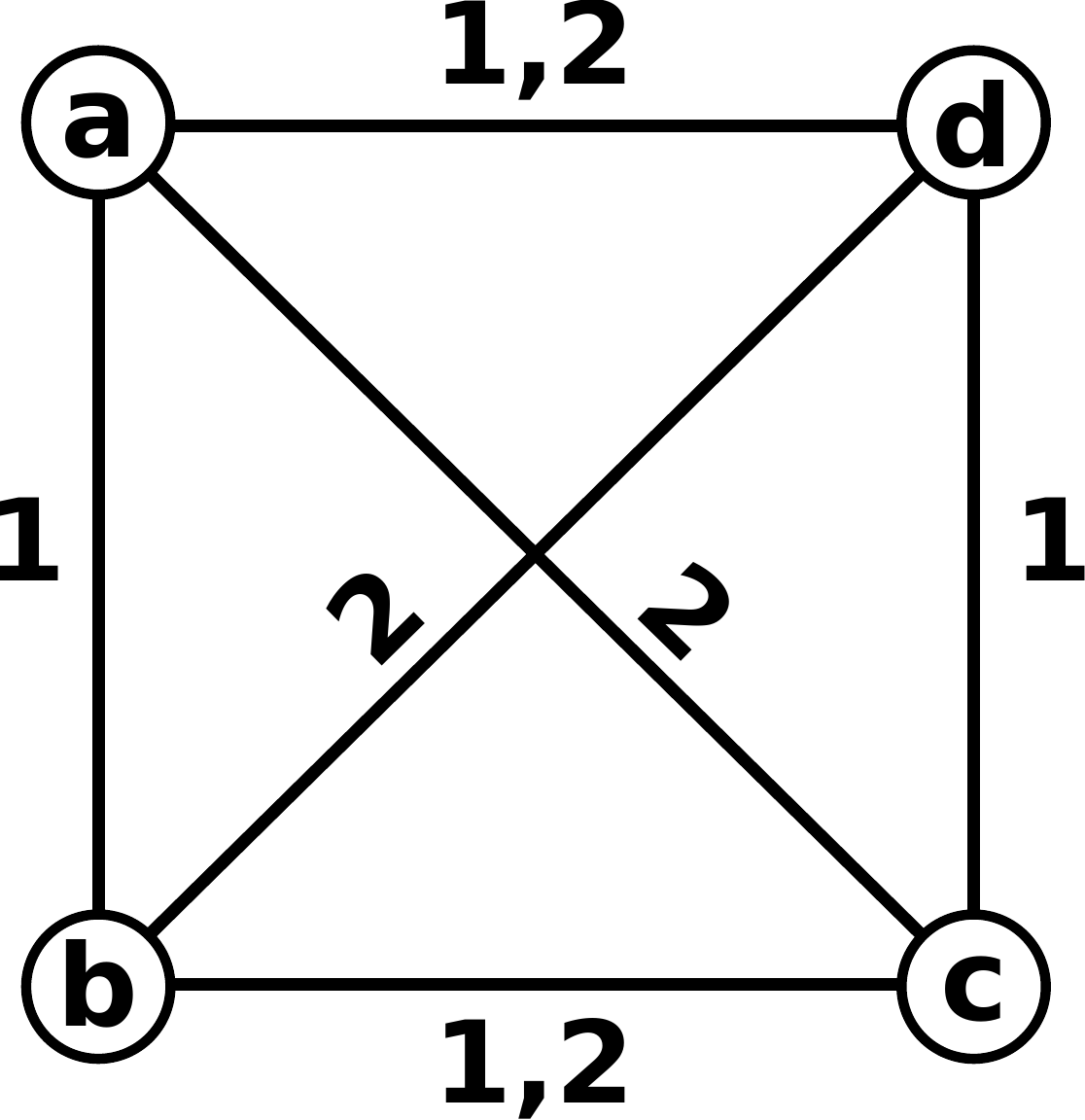}
  \end{center}
  \caption{Let $\eXM$ be a symmetric map with
    $\X\coloneqq \{a,b,c,d\}$ and $M\coloneqq \{1,2\}$, where for every
    distinct $x,y \in \X$ and every $m \in M$ we have $m \in \eps(x,y)$ if
    and only if the edge $\{x,y\}$ in the shown graph contains the label
    $m$.  Then, $\eps$ satisfies \cref{prop:mono-partition}~(1) and (2).
    However, $\eps$ is not a Fitch map, see text for further details. }
	\label{fig:hourglass}
\end{figure}

Unfortunately, the properties in \cref{prop:mono-partition} are not
sufficient to characterize non-monochromatic Fitch maps. To see this,
consider the symmetric map $\eps$ shown in \cref{fig:hourglass}.
Then, we have
$\NnotCol{1}[a]=\NnotCol{1}[c]=\{a,c\}$,
$\NnotCol{1}[b]=\NnotCol{1}[d]=\{b,d\}$,
$\NnotCol{2}[a]=\NnotCol{2}[b]=\{a,b\}$, and
$\NnotCol{2}[c]=\NnotCol{2}[d]=\{c,d\}$.  Hence, both
$\mc{N}\lc{\neg 1}[\eps]=\{ \{a,c\},\{b,d\} \}$ and
$\mc{N}\lc{\neg 2}[\eps]=\{ \{a,b\},\{c,d\} \}$ are partitions of
$\X=\{a,b,c,d\}$.  As we shall prove in \Cref{lem:displ-subsplit} below,
every tree that explains $\eps$ must display the quartets $ab|cd$ and
$ac|bd$.  However, by \Cref{cor:subsplit-equi}, the set $\{ab|cd, ac|bd\}$
of quartets is not compatible. Therefore, $\eps$ cannot be a Fitch map.

Before we provide a characterization of symmetrized Fitch maps, we derive some necessary
conditions.
\begin{lemma} \label{lem:subFitchs} 
	Let $\eXM$ be a symmetrized Fitch map,
  and let $\X'\subseteq \X$ and $M'\subseteq M$.  Then, the map
  $\eps':\irr{\X'\times \X'}\to \PS{M'}$ with
  $\eps'(x,y) \coloneqq \eps(x,y)\cap M'$ for every
  $(x,y) \in \irr{\X'\times \X'}$ is again a symmetrized Fitch map.
\end{lemma}
\begin{proof}
  Let $\eXM$ be a symmetrized Fitch map, and let $\X'\subseteq \X$ and
  $M'\subseteq M$.  Let $\eps':\irr{\X'\times \X'}\to \PS{M'}$ with
  $\eps'(x,y) \coloneqq \eps(x,y)\cap M'$ for every
  $(x,y) \in \irr{\X'\times \X'}$ be a map.
	
  Since $\eXM$ is a symmetrized Fitch map, there is an edge-labeled tree
  $(T,\lambda)$ that explains $\eps$. Now, create a tree $T'$ from $T$,
  where every leaf $x \in \X \setminus \X'$ in $T$ is deleted, and create
  an edge-labeling $\lambda':E(T')\to \PS{M'}$ with
  $\lambda'(e)\coloneqq \lambda(e) \cap M'$ for every $e \in E(T')$.  By
  construction, $m\in \eps'(x,y)$ if and only if the unique path between
  $x$ and $y$ in $T'$ contains an $m$-edge for all $m\in M'$ and
  $x,y\in \X'$. However, the tree $T'$ might have vertices of degree $2$,
  and hence may not be a phylogenetic tree. However, we can further modify
  $T'$ as follows: Suppose that there is a vertex $v$ of degree 2.  Thus,
  there are two edges $e\lc{1}=\{v,w\}$ and $e\lc{2}=\{v,u\}$ in $T'$.
  Now, we remove vertex $v$ and the two edges $e\lc{1}$ and $e\lc{2}$ from
  $T'$ and add the edge $f=\{u,w\}$, and call the resulting tree $T''$.  By
  construction, every path in $T'$ between two leaves $x,y\in \X'$ that
  contains the edge $e\lc{1}$ or $e\lc{2}$ must now contain the edge $f$ in
  $T''$. We construct the edge-labeling $\lambda'':E(T'')\to \PS{M'}$ with
  $\lambda''(e)\coloneqq \lambda'(e)$ for all $e\in E(T'')\setminus f$ and
  $\lambda''(f)\coloneqq \lambda'(e\lc{1})\cup \lambda'(e\lc{2})$.  Then,
  for every $m\in M'$ and every distinct $x,y\in \X'$, we have
  $m \in \eps'(x,y)$ if and only if $m \in \lambda''(e)$ for some edge
  $e \in P\lc{T''}(x,y)$.  Clearly, $T''$ and $\lambda''$ can be
  iteratively modified as described above until no vertices with degree 2
  remain, and hence we end up with an edge-labeled tree
  $(\tilde T, \tilde \lambda)$.  Thus, by construction of $\tilde T$ and
  $\tilde \lambda$, we have $m\in \eps'(x,y)$ if and only if the unique
  path between $x$ and $y$ in $\tilde T$ contains an $m$-edge for all
  $m\in M'$ and $x,y\in \X'$. Hence, $(\tilde T, \tilde \lambda)$ explains
  $\eps'$; and therefore, $\eps'$ is a symmetrized Fitch map.
\end{proof}

\begin{proposition}\label{prop:nec-cond}
  Let $\eXM$ be a symmetrized Fitch map. Then, for every color $m \in M$
  the following \emph{equivalent} statements are satisfied:
  \begin{enumerate}%[noitemsep]
  \item $\G\lc{m}(\eps)$ does not contain a $K_1+K_2$ as an induced subgraph.
    \label{item:k1+k2-graph-forbidden}
  \item For every three pairwise distinct $a,b,c \in \X$ with
    $m \notin \eps(a,b)$ and $m \notin \eps(b,c)$, we have
    $m \notin \eps(a,c)$. \label{item:antitrans}
    \label{item:k1+k2-forbidden}
  \item $\Ns{m}[\eps]$ is a partition of $\X$.
    \label{item:partition}
  \item $\G\lc{m}(\eps)$ is a complete multi-partite graph, where the
    neighborhoods in $\Ns{m}[\eps]$ form precisely the maximal
    independent sets in $\G\lc{m}(\eps)$.
    \label{item:indset}
  \item For every $N \in \Ns{m}[\eps]$, we have $N =\NnotCol{m}[y]$ if and
    only if $y \in N$.
    \label{item:yinN}
  \end{enumerate}
\end{proposition}
\begin{proof}
  Let $\eXM$ be a symmetrized Fitch map, and let $m \in M$ be an arbitrary
  color.  Then, \Cref{lem:subFitchs} implies that the map
  $\eps':\irr{\X\times \X}\to \PS{\{m\}}$ with
  $\eps'(x,y) \coloneqq \eps(x,y)\cap \{m\}$ for every
  $(x,y) \in \irr{\X\times \X}$ is a (monochromatic) symmetrized Fitch
  map. In particular, $\Ns{m}[\eps] = \Ns{m}[\eps']$.  Hence, we can apply
  \Cref{lem:mono-k1+k2,prop:mono-partition} to conclude that the Statements
  \eqref{item:k1+k2-graph-forbidden}, \eqref{item:antitrans} and
  \eqref{item:partition} are satisfied and equivalent.  
  
  We continue by showing the equivalence between Statement
    \eqref{item:partition} and \eqref{item:indset}.  To this end, observe
    first that \Cref{lem:mono-k1+k2}~(1,3) and
    \Cref{prop:mono-partition}~(1,3) directly imply that $\G\lc{m}(\eps)$
    is a complete multi-partite graph if and only if $\Ns{m}[\eps]$ is a
    partition of $\X$. Note, each complete multi-partite graph is, by
    definition, determined by its maximal independent sets.  It remains to
    show that the neighborhoods in $\Ns{m}[\eps]$ are precisely the maximal
    independent sets of $\G\lc{m}(\eps)$. Let
    $\NnotCol{m}[y] \in \Ns{m}[\eps]$.  By definition, for all
    $a,b\in \NnotCol{m}[y]$ we have $m \notin \eps(a,y)$ and
    $m \notin \eps(b,y)$.  Hence, Statement \eqref{item:k1+k2-forbidden}
    implies that $m \notin \eps(a,b)$.  By definition of $\G\lc{m}(\eps)$
    neither of $\{a,y\}$, $\{b,y\}$ and $\{a,b\}$ forms an edge in
    $\G\lc{m}(\eps)$. Hence, $\NnotCol{m}[y]$ is an independent set of
    $\G\lc{m}(\eps)$.  Assume, for contradiction, that $\NnotCol{m}[y]$ is
    not a \emph{maximal} independent set. Hence, there is a vertex
    $z\in V\setminus \NnotCol{m}[y]$ such that
    $\{z,v\}\notin E(\G\lc{m}(\eps))$ for all $v\in \NnotCol{m}[y]$.  In
    particular, therefore, $\{z,y\}\notin E(\G\lc{m}(\eps))$ and thus, by
    definition of $\G\lc{m}(\eps)$, $m\notin \eps(z,y)$.  But then,
    $z\in \NnotCol{m}[y]$; a contradiction.  Therefore, the neighborhoods
    in $\Ns{m}[\eps]$ are precisely the maximal independent sets of
    $\G\lc{m}(\eps)$.

  We continue with showing that Statement \eqref{item:partition} and
  \eqref{item:yinN} are equivalent.  First, suppose that Statement
  \eqref{item:partition} is satisfied, and let $N \in \Ns{m}[\eps]$.  If
  $N=\NnotCol{m}[y]$, then we have by definition $y \in
  \NnotCol{m}[y]=N$. Conversely, if $y \in N$, then we have
  $y \in N\cap \NnotCol{m}[y]\ne \emptyset$.  Hence, since $\Ns{m}[\eps]$
  with $N,\NnotCol{m}[y] \in \Ns{m}[\eps]$ is a partition of $\X$, we
  conclude that $N=\NnotCol{m}[y]$.
	
  Next, we assume that Statement \eqref{item:yinN} is satisfied, and let
  $N,N' \in \Ns{m}[\eps]$ be two arbitrary neighborhoods.  Since we have
  $y \in \NnotCol{m}[y]$ for every $y \in \X$, we conclude that every
  neighborhood is non-empty in $\Ns{m}[\eps]$ and
  $\bigcup_{y \in \X} \NnotCol{m}[y]=\X$.  Moreover, let
  $N\cap N' \ne \emptyset$. Hence, there is a vertex $y \in N\cap N'$, and
  thus by Statement \eqref{item:yinN} we obtain $N=\NnotCol{m}[y]=N'$.  The
  latter arguments together imply that $\Ns{m}[\eps]$ is a partition of
  $\X$, and thus Statement \eqref{item:partition} is satisfied.
\end{proof}

We will need to define certain sets of subsplits associated with the
complementary neighborhoods of $\eps$. 
\begin{definition}\label{def:SsS}
  For a symmetric map $\eXM$ we define the following sets:
  \begin{itemize}%[noitemsep]
  \item The \emph{$m$-subsplit system} of  $\eps$ is
    $\Ss\lc{m}(\eps) \coloneqq \left\{ N|N' \colon N,N'\in
      \Ns{m}[\eps]\text{ and } N\cap N'=\emptyset \right\}$;
  \item The \emph{subsplit system} of  $\eps$ is
    $\Ss(\eps) \coloneqq \bigcup_{m \in M} \Ss\lc{m}(\eps)$; and
  \item the \emph{non-trivial} subsplit system of  $\eps$ is
    $\SsS(\eps) \coloneqq \left\{N|N' \colon N|N' \in \Ss(\eps) \text{ and
      } |N|,|N'|\ge 2 \right\}$.
  \end{itemize}
\end{definition}
Clearly, if a set $\mathcal{S}$ of subsplits is compatible, then every
subset $\mathcal{S}'\subseteq \mathcal{S}$ is also
compatible. $\Ss(\eps)$ is compatible if and only if $\SsS(\eps)$ is
  compatible because  every subsplit
$N|N'\in \Ss(\eps)\setminus \SsS(\eps)$ is trivial and
$\SsS(\eps)\subseteq \Ss(\eps)$.  For later reference we summarize the
latter observation in the following
\begin{lemma} \label{lem:SSSS} Let $\eXM$ be a map. Then, $\Ss(\eps)$ is
  compatible if and only if $\SsS(\eps)$ is compatible.
\end{lemma}

Before we provide our final characterization we observe that compatibility
of $\Ss(\eps)$ is a necessary condition for Fitch maps.
\begin{lemma}\label{lem:displ-subsplit}
  Let $\eXM$ be a symmetrized Fitch map, and let $\Ss(\eps)$ be the
  subsplit system of $\eps$.  Then, every edge-labeled tree $(T,\lambda)$
  that explains $\eps$ displays all subsplits in $\Ss(\eps)$.
\end{lemma}
\begin{proof}
  Let $\eXM$ be a symmetrized Fitch map, and let $(T,\lambda)$ be an
  arbitrary edge-labeled tree that explains $\eps$.  We denote by $T_{|L}$
  the vertex-minimal (not necessarily phylogenetic) subtree of $T$ with
  leaf set $L\subseteq \mc{L}(T)$.

  Assume for contradiction that there is a subsplit $N|N' \in \Ss(\eps)$
  that is not displayed by $T$.  Clearly, if $|N|=1$ or $|N'|=1$, then $T$
  displays $N|N'$. Thus, we can assume that $|N|>1$ and $|N'|>1$.
  Moreover, if none of the paths $P\lc{T}(a,b)$ and $P\lc{T}(c,d)$ with
  $a,b\in N$ and $c,d\in N'$ intersect, then the two trees $T_{|N}$ and
  $T_{|N'}$ are vertex disjoint, and thus, there would be an edge
  $e\in E(T)$ such that $N\subseteq\mc{L}(T_1)$ and
  $N'\subseteq\mc{L}(T_2)$.  Therefore, there are four leaves $a,b \in N$
  and $c,d \in N'$ such that the paths $P\lc{T}(a,b)$ and $P\lc{T}(c,d)$
  intersect.  Hence, there is a vertex
  $v \in V(P\lc{T}(a,b))\cap V(P\lc{T}(c,d))$.
  \Cref{prop:nec-cond}~\eqref{item:yinN}, together with $a,b \in N$ and
  $c,d \in N'$, implies that $a \in N= \NnotCol{m}[b]$ and
  $c \in N'= \NnotCol{m}[d]$. This, together with the fact that
  $(T,\lambda)$ explains $\eps$, implies that there is no $m$-edge on
  either of the paths $P\lc{T}(a,b)$ and $P\lc{T}(c,d)$.  Since $v$ lies on
  both paths $P\lc{T}(a,b)$ and $P\lc{T}(c,d)$, there is no $m$-edge on the
  (sub)paths $P\lc{T}(a,v)$ and $P\lc{T}(v,d)$.  Therefore, the path
  $P\lc{T}(a,d)\subseteq P\lc{T}(a,v)\cup P\lc{T}(v,d)$ cannot contain an
  $m$-edge.  Now, \Cref{prop:nec-cond}~\eqref{item:yinN} and $a\in N$
  imply that $N=\NnotCol{m}[a]$.  However, since $N|N'$ is a subsplit, we
  have $N\cap N'=\emptyset$, and therefore $d \notin N =\NnotCol{m}[a]$.
  This, together with the fact that $(T,\lambda)$ explains $\eps$, implies
  that there is an $m$-edge on the path $P\lc{T}(a,d)$; a contradiction.
  In summary, every subsplit $N|N' \in \Ss(\eps)$ is displayed by $T$.
\end{proof}
	
\Cref{lem:displ-subsplit}, together with \Cref{lem:SSSS}, immediately
implies
\begin{corollary}\label{cor:symFitch-impl-cons}
  If $\eXM$ is a symmetrized Fitch map, then the subsplit sets $\Ss(\eps)$
  and $\SsS(\eps)$ are compatible.
\end{corollary}

\begin{definition}
  \label{def:Teps}
  Let $\eXM$ be a symmetric map such that $\Ss(\eps)$ is
  compatible. Then, we denote with $(\Teps,\leps)$ an edge-labeled tree
  that satisfies the following two conditions:
  \begin{enumerate}%[noitemsep] %,nolistsep]
  \item $\Teps$ displays every subsplit in $\Ss(\eps)$; and
  \item for every edge $e \in E(\Teps)$ we have
    \begin{align*} %\label{equ:def-labeling} \tag{\ref{def:T_eps}a}
      \leps(e) \coloneqq \left\{ m \in M : 
      \begin{array}{l}
        \textnormal{(a) }\, e
        \textnormal{ is a splitting edge w.r.t.\ some }
        N|N' \in \Ss\lc{m}(\eps) \text{ and } \\
        \textnormal{(b) }\, \textnormal{for every } N \in \Ns{m}[\eps]
        \textnormal{ and for every } x,y \in N
        \textnormal{ we have } e \notin E(P\lc{\Teps}(x,y))
      \end{array} 
      \right\}.
    \end{align*}
  \end{enumerate}
\end{definition}
\begin{lemma}\label{lem:sufficiency}
  $\eXM$ is a symmetrized Fitch map if it satisfies the following two
  conditions:
  \begin{enumerate}%[noitemsep]%,nolistsep]
  \item for every $m \in M$ the set $\Ns{m}[\eps]$ forms a partition of
    $\X$; and
  \item $\SsS(\eps)$ is compatible.
  \end{enumerate}
  In particular, $(\Teps,\leps)$ explains $\eps$.
\end{lemma}
\begin{proof}
  Let $\eXM$ be a map that satisfies Conditions~(1) and (2).  Since
  $\SsS(\eps)$ is compatible, \Cref{lem:SSSS} implies that $\Ss(\eps)$ is
  compatible.  Hence, there is a tree $\Teps$ that displays every subsplit
  in $\Ss(\eps)$.  For $\Teps$ let $\leps:E(\Teps)\to\PS{M}$ be the
  edge-labeling as specified in \cref{def:Teps}~(2).  Hence, we obtain an
  edge labeled-tree $(\Teps,\leps)$ that satisfies \cref{def:Teps}.  To
  show that $\eps $ is a symmetrized Fitch map, it suffices to show that
  $(\Teps,\leps)$ explains $\eps$. Thus, we must verify that for every two
  distinct leaves $x,y \in \X$ we have $m \in \eps(x,y)$ if and only if
  there is an $m$-edge on the path $P\lc{\Teps}(x,y)$. To this end, let
  $m \in M$ be an arbitrary color, and let $x,y \in \X$ be two distinct
  arbitrary leaves.

  First, suppose that $m \in \eps(x,y)$. Then, we have
  $y \notin \NnotCol{m}[x]$. This and $y \in\NnotCol{m}[y]$ implies that
  $\NnotCol{m}[x]\ne \NnotCol{m}[y]$. Thus, since $\Ns{m}[\eps]$ is a 
  partition of $\X$, it must hold that
  $\NnotCol{m}[x]\cap \NnotCol{m}[y]=\emptyset$. Therefore, by definition of
  $\Ss(\eps)$, we have
  $\NnotCol{m}[x]|\NnotCol{m}[y] \in
  \Ss\lc{m}(\eps)\subseteq\Ss(\eps)$. Since $\Teps$ displays every subsplit
  in $\Ss(\eps)$, there is a splitting edge $e \in E(\Teps)$ w.r.t.\
  $\NnotCol{m}[x]|\NnotCol{m}[y]$.
  Hence, we have $\NnotCol{m}[x]\subseteq \mc{L}(T_{e,x})$ and
  $\NnotCol{m}[y]\subseteq \mc{L}(T_{e,y})$, where $T_{e,x}$ and $T_{e,y}$
  are the two connected components of $\Teps\setminus e$. We may assume
  w.l.og.\ that this splitting edge $e = \{v,w\}$ w.r.t.\
  $\NnotCol{m}[x]|\NnotCol{m}[y]$ is chosen such that $v$ lies on the
  (unique) path $P\lc{\Teps}(w,x)$ and that $|V(T_{e,x})|$ is minimal among
  all such splitting edges w.r.t.\ $\NnotCol{m}[x]|\NnotCol{m}[y]$.

  There are two cases, either $|V(T_{e,x})|=1$ or $|V(T_{e,x})|>1$.  First,
  suppose that $|V(T_{e,x})|=1$. This is if and only if
  $\mc{L}(T_{e,x}) = V(T_{e,x})=\{v\}=\{x\}$.  Therefore,
    $x\in\NnotCol{m}[x]\subseteq \mc{L}(T_{e,x})=\{x\}$ implies
    $\{x\}=\NnotCol{m}[x]$, and thus, $|\NnotCol{m}[x]|=1$.  Assume for
    contradiction that there is an $N \in \Ns{m}[\eps]$ with $x',y' \in N$
    such that $e \in E(P\lc{\Teps}(x',y'))$.  Since the path
    $P\lc{\Teps}(x',y')$ contains an edge, we conclude that $x'\ne y'$, and
    thus, $|N| \ge 2$.  Therefore, $N \ne \NnotCol{m}[x]$.  Now,
    $N, \NnotCol{m}[x] \in \Ns{m}[\eps]$, which forms a partition of $\X$,
    implies that $N\cap \NnotCol{m}[x]=\emptyset$.  However, since
    $e=\{v,w\}=\{x,w\}$ is an outer edge, we conclude that
    $x \in \{x',y'\} \subseteq N$.  Thus,
    $x \in N \cap \NnotCol{m}[x] \ne \emptyset$; a contradiction.  Hence,
    Condition~(2b) in \cref{def:Teps} is satisfied.  Thus, by construction
  of $\leps$, we have $m\in \leps(e)$. Since $e$ is an edge of the path
  $P\lc{\Teps}(x,y)$ there is an $m$-edge in $P\lc{\Teps}(x,y)$.

  Otherwise, if $|V(T_{e,x})|>1$ and thus $|\mc{L}(T_{e,x})|>1$, then the
  minimality of $|V(T_{e,x})|$ implies that there are two leaves
  $x',x'' \in \NnotCol{m}[x]$ such that $v \in V(P\lc{\Teps}(x',x''))$.

  Now, assume for contradiction that $e$ is not an $m$-edge. Since $e$
  satisfies Condition (2a) in \cref{def:Teps}, it can therefore, not
  satisfy Condition (2b) in \cref{def:Teps}. Hence, there is a neighborhood
  $N' \in \Ns{m}[\eps]$ with $z',z'' \in N'$ such that
  $e \in E(P\lc{\Teps}(z',z''))$. This, together with $e=\{v,w\}$, implies
  $v \in V(P\lc{\Teps}(x',x''))\cap V(P\lc{\Teps}(z',z''))$. Since one of
  the leaves in $\{z',z''\}\subseteq N'$ is not contained in $T_{e,x}$ and
  since $\NnotCol{m}[x]\subseteq\mc{L}(T_{e,x})$, we have
  $N'\neq \NnotCol{m}[x]$.  Since $\Ns{m}[\eps]$ is a partition of $\X$, it
  must hold that $N'\cap\NnotCol{m}[x]=\emptyset$. Therefore,
  $N'|\NnotCol{m}[x]\in \Ss\lc{m}(\eps)\subseteq\Ss(\eps)$. However,
  $v \in V(P\lc{\Teps}(x',x''))\cap V(P\lc{\Teps}(z',z''))$, together with
  $x',x'' \in \NnotCol{m}[x]$ and $z',z'' \in N'$, implies that the
  subsplit $N'|\NnotCol{m}[x] \in \Ss(\eps)$ is not displayed by $\Teps$; a
  contradiction. Therefore, $e$ is an $m$-edge that lies on the path
  $P\lc{\Teps}(x,y)$.
		
  It remains to show that the existence of an $m$-edge on the path
  $P\lc{\Teps}(x,y)$ implies $m \in \eps(x,y)$. Using contraposition,
  assume that $m \notin \eps(x,y)$, and thus
  $x,y \in \NnotCol{m}[x] \in \Ns{m}[\eps]$.  For every edge
  $e \in E(P\lc{\Teps}(x,y))$, Condition (2b) in \cref{def:Teps} is
  violated. Hence, for all $e \in E(P\lc{\Teps}(x,y))$, we have by
  construction of $\leps$ that $m\notin \leps(e)$. Thus, $P\lc{\Teps}(x,y)$
  does not contain an $m$-edge, which completes the proof.

  In summary, we have shown that $(\Teps,\leps)$ explains
  $\eps$. Therefore, $\eps$ is a symmetrized Fitch map.
\end{proof}

The characterization of Fitch maps, which is summarized in
\Cref{thm:charct-part+consist}, follows now directly from
\Cref{prop:nec-cond}~\eqref{item:partition}, \Cref{cor:symFitch-impl-cons}
and \Cref{lem:sufficiency}.
\begin{theorem}\label{thm:charct-part+consist}
  A symmetric map $\eXM$ is a symmetrized Fitch map if and only if
  \begin{enumerate}%[noitemsep]%,nolistsep]
  \item for every $m \in M$ the set $\Ns{m}[\eps]$ forms a partition of
    $\X$; and
  \item $\SsS(\eps)$ is compatible.
  \end{enumerate}
\end{theorem}
For later reference we state here a simple consequence of
\Cref{thm:charct-part+consist}.
\begin{corollary}\label{cor:symFitch-compatible}
  A symmetric map $\eXM$, where $\Ns{m}[\eps]$ forms a partition of
  $\X$ for every $m \in M$, is a symmetrized Fitch map if and only if
  $\SsS(\eps)$ is compatible.
\end{corollary}

By \Cref{prop:nec-cond}, for every symmetrized Fitch map, $\eXM$ the
  graph $\G\lc{m}(\eps)$ is a complete multi-partite graph for each
  $m\in M$, where the neighborhoods in $\Ns{m}[\eps]$ form precisely the
  maximal independent sets in $\G\lc{m}(\eps)$. By \Cref{cor:diam4}, each
  monochromatic symmetrized Fitch map (and thus, the undirected Fitch graph
  $\G\lc{m}(\eps)$) can be explained by unrooted trees $(T_m,\lambda_m)$ of
  diameter $\diam(T_m)\le 4$, see \Cref{fig:least-resolved,fig:diam4} for
  an example. The compatibility problem for $\eps$ is thus related to the
  supertree problem for a collection of trees $T_m$ with diameter at most
  $4$, one for each color $m$. The problems do not appear to be equivalent,
  however, since $\Ns{m}[\eps]$ does not uniquely determine a
  (least-resolved) tree that explains the corresponding Fitch graph.

\section{Complexity Results}

Since monochromatic symmetrized Fitch maps are characterized in terms of
complete multi-partite graphs they can be recognized in polynomial time,
cf.\ \cite{Hellmuth:18a}.  ``Non-symmetrized'' (not necessarily
monochromatic) Fitch maps can also be recognized in polynomial time, cf.\
\cite{Hellmuth2019gf,Hellmuth:2019d}.  However, as we shall show below, the
recognition of symmetrized Fitch maps is \NP-complete, in general.  More
precisely, we consider the following decision problem.
\begin{problem}[\PROBLEM{Symm-Fitch Recognition}]\ \\
  \begin{tabular}{ll}
    \emph{Input:}    & A symmetric map $\eXM$.  \\
    \emph{Question:} & Is $\eps$ a symmetrized Fitch map, i.e., is
                       there an edge-labeled tree $(T,\lambda)$
                       that explains $\eps$?
  \end{tabular}
\end{problem}
In order to prove \NP-completeness, we use a reduction from the following
\NP-complete problem \cite{Steel1992}.
\begin{problem}[\PROBLEM{Quartet Compatibility}]\ \\
  \begin{tabular}{ll}
    \emph{Input:}    & A set $Q$ of quartets on \X.  \\
    \emph{Question:} & Is $Q$ compatible?
  \end{tabular}
\end{problem}
\begin{proposition}[{\cite[Thm.\ 1]{Steel1992}}]\label{thm:quartet-NP-hard}
  \PROBLEM{Quartet Compatibility} is \NP-complete.
\end{proposition}

\begin{theorem}\label{thm:symFitch-NP-hard}
  \PROBLEM{Symm-Fitch Recognition} is \NP-complete.
\end{theorem}
\begin{proof}
  Clearly, \PROBLEM{Symm-Fitch Recognition} $\in$ \NP, since we can test in
  polynomial time whether a given edge-labeled tree $(T,\lambda)$ indeed
  explains $\eps$.

  Let $\mathcal{Q}=\{q\lc{1},q\lc{2},\ldots,q\lc{|Q|}\}$ be an arbitrary
  set of quartets on $\X$.  Now, we construct a map $\eXM$ with
  $M=\{1,2,\ldots,|\mathcal{Q}|\}$ such that for every
  $(x,y) \in \irr{\X\times \X}$ we have
  \begin{align*}
    \eps(x,y)\coloneqq
    \Big\{ i \in M : q\lc{i}=ab|cd \textnormal{ and }
    \{x,y\}\notin\big\{ \{a,b\},\{c,d\} \big\}  \Big\}.
  \end{align*}
  By construction of $\eps$ we have for every $q_i=ab|cd \in \mathcal{Q}$:
  \begin{align*}
    \NnotCol{i}[a]&=\NnotCol{i}[b]=\{a,b\}, \\
    \NnotCol{i}[c]&=\NnotCol{i}[d]=\{c,d\}, \textnormal{ and} \\
    \NnotCol{i}[y]&=\{y\} \textnormal{ for every } y \in \X\setminus\{a,b,c,d\}.
  \end{align*}
  Hence, $\Ns{i}[\eps]$ is a partition of $\X$ for every color $i \in M$.
  Now, we continue to show that $\mathcal{Q}=\SsS(\eps)$.  If
  $q_i=ab|cd \in \mathcal{Q}$ then, by construction of $\eps$, we have
  $ab|cd=\NnotCol{i}[a]|\NnotCol{i}[c] \in \SsS(\eps)$.  Conversely, if
  $ab|cd \in \SsS(\eps)$, then there is a color $i \in M$ such that
  $\NnotCol{i}[a]=\{a,b\}$ and $\NnotCol{i}[c]=\{c,d\}$.  This and the
  construction of $\eps$ imply that $ab|cd=q_i\in \mathcal{Q}$.  Thus, we
  have $\mathcal{Q}=\SsS(\eps)$.  Since $\Ns{i}[\eps]$ is a partition of
  $\X$ for every color $i\in M$, we can apply
  \cref{cor:symFitch-compatible} to conclude that $\eps$ is a symmetrized
  Fitch map if and only if $\SsS(\eps)=\mathcal{Q}$ is compatible.
	
  Since deciding whether $\mathcal{Q}$ is compatible is \NP-complete, see
  \cref{thm:quartet-NP-hard}, we can conclude that deciding whether $\eps$
  is a symmetrized Fitch map is \NP-hard.  This, together with
  \PROBLEM{Symm-Fitch Recognition} $\in$ \NP, implies that
  \PROBLEM{Symm-Fitch Recognition} is \NP-complete.
\end{proof}

We note in passing that \Cref{thm:symFitch-NP-hard} implies that there is
no characterization of Fitch maps in terms of a finite set of forbidden
subgraphs (unless $\textsf{P}=\NP$).

\section{Summary and Outlook}

In this contribution, we have characterized a class of symmetric maps
$\eXM$, or equivalently, sets of (not necessarily disjoint) symmetric
binary relations $R_1,\dots R_{|M|}$ that arise in a natural way from
edge-labeled trees with a set of ``colors''. The symmetrized Fitch maps are
those for which $\eps(x,y)$ is the set of colors encountered along the
unique path connecting $x$ and $y$ in $T$. In the monochromatic cases
$|M|=1$ there is only a single relation $R_1$ (or graph). As already shown
by \cite{Hellmuth:18a}, $R_1$ is symmetrized Fitch relation if and only if
it is a complete multi-partite graph. Here we provide an alternative
characterization in terms of complementary neighborhoods. Restricted
symmetrized Fitch maps assign at most one color to each pair $(x,y)$, i.e.,
$|\eps(x,y)|\le 1$. We found that these two classes coincide. Therefore,
such maps can be recognized in polynomial time.  In the general case, we
obtained a series of necessary conditions as well as a characterization in
terms of monochromatic ``induced'' submaps and certain subsplits defined by
the complementary neighborhoods of $\eps$ that must be displayed by every
tree explaining $\eps$, i.e., the subsplit system must be compatible. These
result were utilized to show that the recognition of symmetrized Fitch maps
is \NP-complete.

Although we have obtained a comprehensive characterization interesting open
questions remain. The complete multi-partite graphs are a subclass of the
cographs, i.e., graphs that do not contain a path of length four as in
induced subgraph \cite{Corneil:81,Corneil:85}.  Cographs can be explained
by vertex-labeled trees.  In particular, the di-cograph structure
\cite{Crespelle:06} of non-symmetrized Fitch maps has been very helpful in
the construction of efficient recognition algorithms \cite{Geiss:18a} for
the directed case. Since for every color $m\in M$ the graph-representation
$\G\lc{m}(\eps)$ of a symmetrized Fitch map $\eps$ must be a complete
multi-partite graph, $\G\lc{m}(\eps)$ is a cograph.  Clearly, this does not
help directly for efficient recognition algorithms since the recognition
problem is \NP-complete. However, if we restrict our attention to maps
$\eXM$ that additionally satisfy the ``triangle condition''
$|\{\eps(x,y), \eps(x,z), \eps(y,z)\}|\leq 2$ for every pairwise distinct
$x,y,z \in \X$, then we obtain the subclass of so-called \emph{unp
  2-structures} \cite{Hellmuth:17a}, which can be recognized in polynomial
time. In future work we will investigate whether symmetrized Fitch map that
satisfy this triangle condition can be recognized in polynomial time.
Complementary, one may ask whether there are interesting constellations of
complementary neighborhoods for which compatibility of $\SsS(\eps)$ can be
checked efficiently, e.g.\ by the All Quartets Algorithm \cite[Sect.\
5.2]{Warnow:17}.

In \cite{Hellmuth2019gf}, we characterized non-symmetrized
``$k$-restricted'' Fitch maps that can be explained by edge-labeled trees
$(T,\lambda)$ with $|\lambda(e)|\leq k$ for every $e \in E(T)$ and some
fixed integer $k$. This characterization was entirely based on the
cardinality of complementary neighborhoods and the proof relied on
the fact that the least-resolved tree for a non-symmetrized Fitch map is
unique.  However, finding a characterization for ``$k$-restricted''
symmetrized Fitch maps, seems to be quite difficult, since we cannot build
upon the fact that least-resolved trees are unique for symmetrized Fitch
maps (see \cref{fig:least-resolved} for a counterexample).  Thus, it
remains an open question if such restrictions may lead to deeper
understanding of symmetrized Fitch maps and whether such maps can be
recognized in polynomial time or not.

Real-life estimates of graphs are usually subject to measurement
errors. Attempts to correct these estimates naturally leads to editing
problem.  In our setting, given a symmetric map $\eps$, we are interested
in a symmetrized Fitch map $\eps'$ that is ``as close as possible'' to
$\eps$. A natural distance measure is e.g.\ the sum of the symmetric
differences of the edges of $\G\lc{m}(\eps)$. In the light of
\Cref{cor:symFitch-compatible} one may ask whether there is a connection
between this ``Fitch Map Editing'' problem and the problem of finding a
maximal subset of consistent quartets in $\SsS(\eps)$. Conversely, can one
of the many heuristics for the \textsc{Maximum Quartet Consistency Problem}
(see \cite{Morgado:10,Reaz:14} and the references therein) be adapted such
that $\Ns{m}[\eps]$ remains a partition for every $m \in M$?

\acknowledgements
We thank the anonymous referees for their constructive comments that helped
to significantly improve the paper. In particular, their comments opened
the avenue to characterize least- and minimally-resolved trees for 
monochromatic symmetrized Fitch relations.

\nocite{*}
\bibliographystyle{abbrvnat}
% use the following instead if you encounter problems 
%\bibliographystyle{alpha}
\bibliography{GenFitch-sym}
\label{sec:biblio}

\end{document}